\documentclass[12pt,aps,prd,showpacs,raggedbottom,nobalancelastpage,amssymb,superscriptaddress,longbibliography]{revtex4-1}

\usepackage{amsmath, amssymb, slashed,graphicx,color}
\usepackage[toc,page]{appendix}
\usepackage[colorlinks=true,urlcolor=black,linkcolor=red,citecolor=blue]{hyperref}
\usepackage[margin=1in]{geometry}
\usepackage[utf8]{inputenc}

\newcommand{\balpha}{{\boldsymbol{\alpha}}}
\newcommand{\bbeta}{{\boldsymbol{\beta}}}
\newcommand{\bsigma}{{\boldsymbol{\sigma}}}
\newcommand{\bgamma}{{\boldsymbol{\gamma}}}
\newcommand{\blambda}{{\boldsymbol{\lambda}}}
\newcommand{\bmu}{{\boldsymbol{\mu}}}
\newcommand{\bnu}{{\boldsymbol{\nu}}}

\newcommand{\Gamme}{\mathsf{\Gamma}}

\newcommand{\uw}{\widetilde{W}}
\newcommand{\uf}{\widetilde{\Gamme}}
\newcommand{\D}{\mathsf{D}}
\newcommand{\G}{\mathsf{G}}
\newcommand{\h}{\mathsf{H}}
\newcommand{\M}{\mathcal{M}}

\newcommand{\oxi}{\overline{\xi}}
\newcommand{\vphi}{\varphi}
\newcommand{\Tr}{{\rm Tr} }

\usepackage{ulem}

\begin{document}

\title{Non-linear sigma models on constant curvature target manifolds: a functional renormalization group approach}

\author{Alexander N. Efremov}
\affiliation{Universit\'e  de  Lille,  CNRS,  UMR  8523  --  PhLAM  --  Laboratoire  de Physique  des  Lasers,  Atomes  et  Mol\'ecules,  F-59000  Lille,  France}
\author{Adam Ran\c{c}on}
\affiliation{Universit\'e  de  Lille,  CNRS,  UMR  8523  --  PhLAM  --  Laboratoire  de Physique  des  Lasers,  Atomes  et  Mol\'ecules,  F-59000  Lille,  France}

\begin{abstract}
We study non-linear sigma models on target manifolds with constant (positive or negative) curvature using the functional renormalization group and the background field method. We pay particular attention to the splitting Ward identities associated to the invariance under reparametrization of the background field. Implementing these Ward identities imposes to use the curvature as a formal expansion parameter, which allows us to close the flow equation of the (scale-dependent) effective action consistently to first order in the curvature. We shed new light on previous work using the background field method.
\end{abstract}

\date{\today}
\maketitle

\section{Introduction}
The non-linear  sigma models (NLSM) are a very rich class of dynamical systems which spans many fields of physics such as for example high energy physics, string theory, statistical physics. For instance, the $O(4)$ NLSM first appeared in the work of M. Gell-Mann and M.  Lévy as an effective model of pion-nucleon interaction~\cite{gellmann}. More recently L.D. Faddeev has shown that the $O(3)$ NLSM with a topological term might appear in the confined phase of the $SU(2)$ Yang--Mills theory if one performs the Spin-Charge decomposition~\cite{fad1}. However there is no any proof of existence of the quantum model at the present time. 
NLSM on a two dimensional manifold, i.e. the string world sheet, appear in string theory~\cite{hull}. In general relativity one can consider the metric tensor as a Goldstone boson identified with the coset $GL(4,\mathbb{R})/SO(1,3)$~\cite{percacci2}. It is therefore a NLSM which is similar to the Skyrme model. Furthermore one is often interested in the asymptotic safety of this sigma-model in more than two dimensions. In the language of Wilson's renormalization group a theory is asymptotically safe if the critical surface has a finite co-dimension, i.e. Weinberg's ultraviolet critical surface is finite dimensional~\cite{weinberg}.

In statistical physics, NLSM are used to describe spin systems, especially close to two dimensions \cite{Polyakov1975,Brezin1976}. In this context, it is widely believed that the $O(N)$ NLSM belongs to the same universality class than the $O(N)$ linear sigma model (a $\phi^4$ theory), which has a non-trivial infrared fixed point only in spatial dimensions $2<d<4$ (we only refer to the case $N>2$ for simplicity). This fixed point describes the critical state between an ordered phase (described to a weak coupling fixed point in the NLSM) and 
and a disordered phase (with massive modes, corresponding to a strongly coupled theory). Importantly, this non-trivial Wilson-Fisher fixed point merges with the Gaussian  fixed point (in the context of the linear model) at the upper critical dimension $d_c=4$, meaning that critical behavior is captured by a free field theory.

The NLSM on non-compact target-manifolds are also relevant to the physics of Anderson localization \cite{Schaefer1980,Houghton1980}, see \cite{Evers2008} for a review. In this context, a toy model corresponds to the target space $SO(1,N-1)/O(N-1)$, in the ``replica limit'' $N\to 1$ \cite{Gruzberg1996} (see also \cite{Cohen1983,Amit1983,Niedermaier2008,Niedermaier2008a} for applications of this model in high-energy physics). The physics is expected to be rather different from that of its compact counterpart. In particular, it is believed that the upper critical dimension is infinite, with non-trivial critical exponents in all dimensions $d>2$, see \cite{Tarquini2017} for a recent analysis of this issue.

Wilson's Renormalization Group (RG) is the method of choice to address phase transitions, developed originally in statistical mechanics~\cite{Wilson1974,wegner} and later extended to the field theory~\cite{pol84}.  An application of these ideas to the non-linear $\sigma$-model follows one of two directions. In the first approach one considers the linear $\sigma$-model with an axillary non-linear constraint~\cite{mitter89}. The second method is based on the covariant Taylor expansion around a background field~\cite{honerkamp72} and is more natural for the $\sigma$-models which are not multiplicatively renormalizable in general. 

The Functional RG (FRG) is a modern implementation of the RG which allows for non-perturbative approximations, see \cite{Dupuis2021} for a recent review. The background field method has been adapted to the FRG for applications in quantum gravity and non-Abelian gauge theories. Note that the invariance under reparametrization of the background field give rise to the so-called splitting Ward identities \cite{howe88,safari16}. Somewhat surprisingly, the FRG with the background field method has only been used quite recently to study the $O(N)$ NLSM \cite{percacci,flore,Flore2013a}. In  \cite{percacci}, the flow equation at lowest order in the derivative expansion were obtained, and a non-trivial fixed point is found for all $d>2$. In \cite{flore}, the expansion is pushed to the next order, and the fixed point seems to disappear if all the allowed coupling constants are kept in $d=3$, where the existence of a non-trivial fixed point is beyond doubt. In all these previous studies the authors have not taken into account the splitting Ward identities to organize their approximations.

Here, we revisit this problem, generalizing the analysis to arbitrary constant curvature target-manifold. We use the lowest order in the derivative expansion, as in \cite{percacci}, but implement the splitting Ward identities explicitly, which leads to different flow equations and emphasizes the importance of the Ward identities in the background field method for the non-linear $\sigma$-model. Unfortunately, the splitting Ward identity can only be written in terms of a formal expansion in the curvature of the target-manifold, and our flow equations are therefore only valid to lowest order in derivatives and curvature.

The manuscript is organized as follows.
In section~\ref{sec_model} we define the model and review the background field method, and describe the FRG in Sec.~\ref{sec_FRG}. In section~\ref{sec_WI} we give the splitting symmetry transformation for the symmetric manifolds up to second order in the Riemann tensor.  An explicit form of the transformation is needed to impose the splitting Ward identities~\cite{howe88,safari16} for the FRG. As a consequence the curvature becomes the main expansion parameter in our work. The flow equations at lowest order in the curvature are derived in Section~\ref{sec_flow}. In particular we find that the coupling constants have different evolution equations, in contrast to what happens in linear models. Using the Ward identities, we are able to close the flow equations, the corresponding beta functions are given in Sec.~\ref{sec_betafunc}. We discuss our results, and compare them to previous studies using the background field method, in Sec.~\ref{sec_concl}.

\section{Model and background-field expansion}\label{sec_model} 
 For $\M$ a simply connected $\D$-dimensional ($\D=N-1$) manifold of a constant curvature $K$ endowed with a  metric~$h$, the  Levy--Civita connection~$D$ compatible with the metric~$D h=0$; $\phi^{-1}:  \M \to \mathbb{R}^d$ a chart on $\M$, the action of the NLsM on the target space $\M$ is defined as
\begin{equation}
  S(\phi)=  \frac{1}{2 t}\int_x h_{\alpha \beta}(\phi) \, \partial_i \phi^\alpha  \partial_i \phi^\beta ,
\end{equation}
where $\partial_i \phi^\alpha=\frac{\partial }{\partial x^i}\phi^\alpha(x)$, $\alpha=1 \dots \D$, $i = 1 \dots d$, and $\int_x\equiv \int d^d x$.   Here $t>0$ is a nonperturbative bare coupling constant which is proportional to the temperature in the Heisenberg model~\cite{Brezin1976}. The theory is regularized by an ultraviolet cut-off $\Lambda$. 
For a positive curvature manifold $h$ is an elliptic metric. For a negative curvature manifold $h$ is a hyperbolic metric. In the both cases $h$ is positive definite. In the following, we will consider the case of a $\D$-dimensional sphere~$\M=\mathbb{S}^\D$  for $K>0$ and  the hyperbolic space~$\mathbb{H}^{\D}$ for $K<0$.

We consider the functional RG in the context of the covariant background-field method, i.e. by writing the field $\phi(x)$ in terms of a (fixed) background $\varphi(x)$ and the corresponding (fluctuating) normal  field $\xi(x)\in T_{\varphi(x)} \M$ \cite{honerkamp72}. Assume that there is a smooth map $\phi_s(x)$ such that $\phi_0(x)=\varphi(x)$ and $\phi_1(x)=\phi(x)$ with $\dot{\phi}_0=\xi$.
We choose the curve $\phi_s$ in $\M$ to coincide with the geodesic between the initial and final points $\phi_0$, $\phi_1$, i.e.
\begin{align}
  \ddot{\phi}^{\balpha}_s + \Gamma^{\balpha}_{\bsigma \bgamma} \dot{\phi}^{\bsigma}_s \dot{\phi}^{\bgamma}_s&=0,&\Gamma^{\alpha x}_{\sigma z  \; \gamma z^\prime}&=\Gamma^{\alpha}_{\sigma \gamma}(\phi_x) \delta_{x  z} \delta_{x z^\prime},
\end{align}
where $\Gamma^{\alpha}_{\sigma \gamma}(\phi)$ is the Christoffel symbol and $\balpha$ stands for the multi-index $(\alpha,x_1,\dots, x_d)$.
Then for a smooth functional~$f$ we have the Taylor expansion
\begin{equation}
  f(\phi)=f(\phi_0)+ \dot{\phi}^\balpha_s \frac{\delta f(\phi)}{\delta \phi^\balpha}\Big|_{s=0} + \frac{1}{2} \left( \ddot{\phi}^{\balpha}_s \frac{\delta f(\phi)}{\delta \phi^\balpha} + \dot{\phi}^\balpha_s\dot{\phi}^\bbeta_s \frac{\delta^2 f(\phi)}{\delta \phi^\balpha \delta \phi^\bbeta} \right)\Big|_{s=0}+ \dots,
\end{equation}
 which can be written as 
\begin{align}
  f(\phi)&=e^{\xi^\balpha D_\balpha} f(\varphi),&D_{\balpha} f^{\bbeta}(\varphi)= \frac{\delta f^\bbeta(\varphi)}{\delta \varphi^\balpha} + \Gamma^{\bbeta}_{\bsigma \balpha} f^{\bsigma}(\varphi).
\end{align}
The functional $f[\vphi,\xi]=f[\phi(\vphi,\xi)]$ depends only on $\phi$ and not on the way the splitting between $\vphi$ and $\xi$ is done. In other words for $\tilde \varphi$ a new expansion point and $\tilde \xi$ the corresponding new normal field such that $\phi(\vphi,\xi)=\phi(\tilde\vphi,\tilde\xi)$ we still have $f[\vphi,\xi]=f[\tilde\vphi,\tilde\xi]$. Such a functional is called a ``single field'' functional \cite{flore}. This invariance will impose strong constraints on the FRG functionals, as discussed in Sec.~\ref{sec_WI}. To calculate the expansion coefficients one uses the standard relations
\begin{align}
 &\xi^\balpha D_\balpha \partial_i \varphi^\blambda = \partial_i \xi^\blambda   + \Gamma^{\blambda}_{\balpha  \bgamma } \xi^\balpha \partial_i \varphi^{\bgamma} =D_i \xi^{\blambda},\\
  &D_\balpha D_\bbeta \partial_i \varphi^\blambda=  R^\blambda_{\bbeta  \balpha \bgamma} \partial_i \varphi^\bgamma,
\end{align}
where $R^{\blambda}_{\bbeta \balpha \bsigma}$ is the Riemann tensor,
\begin{equation}
   R^{\blambda}_{\bbeta \balpha \bsigma}(\varphi)= \frac{\delta \Gamma^{\blambda}_{ \bbeta \bsigma}}{ \delta \varphi^{\balpha} } - \frac{\delta \Gamma^{\blambda}_{ \bbeta \balpha}}{ \delta \varphi^{\bsigma} } + \Gamma^{\blambda}_{\bgamma \balpha} \Gamma^{\bgamma}_{ \bbeta \bsigma} - \Gamma^{\blambda}_{\bgamma \bsigma} \Gamma^{\bgamma}_{\bbeta \balpha}.
\end{equation} 
Note that $R^{\blambda}_{\bbeta \balpha \bsigma}(\varphi)$ is ultra-local, i.e. it is proportional to the product of three delta functions of the space coordinates.
Furthermore for a constant curvature manifold we have
\begin{align}
  R_{\alpha \lambda \beta  \gamma}=&K \Pi_{\alpha  \lambda \beta \gamma} ,& \Pi_{\alpha  \lambda \beta \gamma}&=h_{\alpha \beta} h_{\lambda \gamma}- h_{\alpha \gamma} h_{\lambda \beta }.
\end{align}
 Here and below $h_{\alpha \beta}=h_{\alpha \beta}(\varphi)$ denotes the metric tensor at the expansion point. Clearly the curvature tensor is covariantly constant $D_\bgamma R^{\blambda}_{\bbeta \balpha \bsigma}=0$.

\section{Functional RG \label{sec_FRG}}

The strategy of the FRG is to build a family of models, indexed by a momentum scale $k$, which interpolates between the semiclassical limit for $k = \Lambda$ and the model of interest for $k \to 0$. For this purpose, one introduces a regulator term $\Delta S_k$ in the action, which leaves the modes with momentum larger than $k$ untouched while freezing the low-momentum modes, implementing effectively Wilson's RG.

We first introduce the generating functional of n-point connected Schwinger functions~$W_k[\vphi,j]$, which depends on the background $\vphi$ and source $j\in T_\varphi \M$ linearly coupled to the normal field~$\xi$~\cite{vil84,hull86,burgess87},
\begin{equation}
e^{W_k[\vphi,j]}=\int \mathcal D_\vphi (\xi) e^{-S[\varphi, \xi]-\Delta S_k[\vphi,\xi]+j.\xi}.\label{2107a}
\end{equation}
For details see App.~\ref{2107c}. The measure
\begin{equation}
\mathcal D_\vphi( \xi)= ({\rm Det}- \partial^2_\Lambda)^{\frac{\D}{2}}\sqrt{{\rm Det} \, h_\Lambda}\, e^{-U_\Lambda[\vphi,\xi]}\prod \limits_{\balpha}\frac{d \xi^\balpha}{\sqrt{2 \pi}} 
\label{eq_Dxi}
\end{equation}
corresponds to the invariant measure after the change of variables from $\phi$ to $\xi$ at fixed background. It has been convenient to introduce $U_\Lambda[\vphi,\xi]=-\log\left(\sqrt{{\rm Det} \, h^{-1}_\Lambda h_\Lambda(\phi) } \left|\frac{\delta\phi}{\delta\xi}\right|\right)$, corresponding to an ultra-local term in the action which can be expanded in $\xi$. It is necessary to include this term to preserve the symmetries of the background expansion explicitly, see Sec.~\ref{sec_WI}. 
 This term contains the Dirac delta at zero $\delta_0$ and thus it is meaningful only in the presence of the ultraviolet regularization. Introducing new constants $\rho_{i, \Lambda}=\delta_0$, the expansion of~$U_\Lambda$ in the normal fields~$\xi$ reads \cite{muller99}
\begin{equation}
U_\Lambda[\varphi,\xi] = \rho_{2,\Lambda} U^{(2)}[\varphi,\xi]   + \rho_{4,\Lambda} U^{(4)}[\varphi,\xi]    + o(\xi^4),
\end{equation}
where
\begin{equation}
\begin{split}
U^{(2)}[\varphi,\xi]  &=\frac{1}{6} R_{\alpha \beta} \int_x \xi^\alpha \xi^\beta=   \frac{K(\D -1)}{6} \int_x  \xi_\alpha\xi^\alpha,\\
U^{(4)}[\varphi,\xi]  &= \frac{1}{180} R^\sigma_{\alpha \beta \gamma} R^\gamma_{\mu \nu \sigma} \int_x\xi^\mu \xi^\nu   \xi^\alpha \xi^\beta =  \frac{K^2 (\D-1)}{180}  \int _x( \xi_\alpha\xi^\alpha )^2.
\end{split}
\label{eq_U}
\end{equation}
In perturbation theory, one usually uses dimensional regularization, for which the $\rho_{i,\Lambda}$ vanish. In contrast, in the FRG, one works (sometimes implicitly) with a momentum cut-off $\Lambda$, implying a non-zero $\rho_{i,\Lambda}$, as was done in particular in the FRG study of the NLsMs for example in~\cite{percacci,flore} (see however \cite{Baldazzi2021} for an attempt to reproduce the $\beta$-function in the $\overline{MS}$ scheme with FRG).  

For later convenience, we give the expansion of the action to quadratic order in $\xi$~\cite{mukhi81,howe88,ket2000}
\begin{equation}
\begin{split}
 S[\phi]=&\frac{1}{2 t}\int_x h_{\alpha \beta}(\vphi) \, \partial_i \vphi^\alpha  \partial_i \vphi^\beta- \frac{1}{t}  \int_x \xi_\alpha D_i \partial_i \varphi^\alpha\\
  & +\frac{1}{ 2t} \int_x  \xi^\alpha \left(-  h_{\alpha\beta}D^2  +E_{\alpha \beta}\right)\xi^\beta + o(\xi^2),
\end{split}
  \label{eq_action_exp}
\end{equation}
with $E_{\alpha \beta}=- K \Pi_{\alpha \lambda \beta \gamma}  \partial_i \varphi^\lambda \partial_i \varphi^\gamma$. More terms are given in App.~\ref{app_action}.

Contrary to the action, the regulator term 
$\Delta S_k[\vphi,\xi]=\frac12 \xi^\balpha \mathcal{R}_{\balpha \bbeta,k}[\vphi]\xi^{\bbeta}$  is a ``two-field'' functional as it depends independently on the fields $\vphi$ and $\xi$, and cannot be written as a functional of $\phi$ only. 

Introducing the classical fields $\oxi_\balpha=\langle\xi_\balpha\rangle=\frac{\delta W_k}{\delta j_\alpha}$, the scale-dependent Wetterich's effective action is defined as a modified Legendre transform of $W_k$,
\begin{equation}
\Gamma_k[\vphi,\oxi]=-W_k[\vphi,j]+j.\oxi-\Delta S_k[\vphi,\oxi].\label{2107b}
\end{equation}
The assumption that $\mathcal{R}_{\Lambda}=\infty$, see e.g. App.~\ref{2107c}, gives the initial condition in the form 
\begin{equation}
\lim \limits_{k \to \Lambda} \left(\Gamma_{k, \varphi}[\vphi,\oxi] - \frac{1}{2} \Tr \log (-\partial^{2}_{\Lambda})^{-1}  (- D^{2}_{\Lambda}+ \mathcal{R}_{k}) \right) =S_\Lambda[\varphi,\oxi]+U_\Lambda[\varphi,\oxi].
\end{equation}
We use in practice a regulator~$\mathcal{R}_k$ which is finite at the boundary, i.e. $\mathcal{R}_{\Lambda}\propto \Lambda^2$. Since we are only interested in the behaviour of the RG flow near fixed points, we will keep the original boundary conditions unchanged and instead consider the effective action at $k=\Lambda$ as a perturbation of the semiclassical model. It is believed that the trajectory of the perturbed system on the phase diagram will remain within a small distance from the trajectory of the model. Since $\mathcal{R}_{0}=0$, the functional $\Gamma_{k=0}[\vphi,\oxi]$ coincides with the Vilkovisky--Dewitt effective action~\cite{vil84}.

The scale-dependent effective action obeys the exact RG equation \cite{Wetterich1993}
\begin{equation}
    \partial_k \Gamma_k[\vphi,\oxi]= \frac{1}{2} \Tr\left( \partial_k \mathcal{R}_k \left(\Gamma^{(2)}_k + \mathcal{R}_k \right)^{-1} \right), 
    \label{eq_Wett}
\end{equation}
where the trace is over space and the internal degrees of freedom. Here and below we use the following notation
\begin{equation}
  \Gamma^{(n)}_{\balpha_1\ldots\balpha_n,k}[\vphi,\oxi]=\frac{\delta^n\Gamma_k}{\delta \oxi^{\balpha_1}\ldots\delta \oxi^{\balpha_n}}\;.
  \end{equation}
The exact flow equation is difficult to solve. Since we are interested in $\Gamma_k[\vphi,0]$ and in the long-distance physics, it is natural to restrict the effective action to a subspace of functionals with a fixed number of derivatives. However the normal fields~$\oxi$ are dimensionless and this truncation is not enough to obtain a finite dimensional dynamical system. In the background field method one usually retains only the evolution equation for the background action $\Gamma_k[\varphi,0]$ omitting the equations corresponding to $n$-point vertex functions. To close the obtained dynamical system, we will rely on the splitting Ward identities associated with the splitting into the background and fluctuation fields. Since these Ward identities are a formal series in the curvature $K$, we will use $K$ as the main expansion parameter. 
 
 To leading order in $K$ and in derivatives, we use the following ansatz
\begin{equation}
\begin{split}
  \Gamma_k[\vphi,\oxi]=&\frac{1}{2 t_{0,k}}\int_x h_{\alpha \beta}(\vphi) \, \partial_i \vphi^\alpha  \partial_i \vphi^\beta- \frac{1}{t_{1,k}}  \int_x \oxi_\alpha D_i \partial_i \varphi^\alpha\\
  & + \frac12\int_x  \oxi^\alpha \left(-  \frac{1}{ t_{2,k}}h_{\alpha\beta}D^2  +\upsilon_k E_{\alpha \beta}+w_k E^\gamma_{\; \gamma}h_{\alpha\beta}\right)\oxi^\beta   + V_k[\varphi,\oxi] + U_k[\vphi,\oxi],
\end{split}
\label{eq_ansatz}
\end{equation}
where to lowest order in $K$, we can use
\begin{equation}
\begin{split}
V_k[\varphi,\oxi] &= \frac{1}{t_{3,k}} V^{(3)}[\varphi,\oxi]+\frac{1}{t_{4,k}} V^{(4)}[\varphi,\oxi]+o(\xi^4),\\
U_k[\varphi,\oxi] &= \rho_{2,k} U^{(2)}[\varphi,\oxi]+ \rho_{4,k}U^{(4)}[\varphi,\oxi]+o(\xi^4).
\end{split}
\label{eq_VU}
\end{equation}
For~$V^{(n)}[\varphi,\oxi]$ see App.~\ref{app_action}, the functional $U^{(n)}[\varphi,\oxi]$ is given in Eq.~\eqref{eq_U}. All other terms generated by the renormalization flow contribute to the second order in~$K$. This is why we do not include them in the truncation.

Comparing to the covariant Taylor expansion of the action, Eq.~\eqref{eq_action_exp}, we find the initial conditions
\begin{equation}
\begin{split}
t_{i,\Lambda}=t,\\
\rho_{i,\Lambda}=\delta_0,\\
\upsilon_\Lambda=t^{-1},\\
w_\Lambda=0.
\end{split}
\label{eq_IC}
\end{equation}
Note that while all $t_{i,k}$ are equal at the beginning of the flow, this is not so for all $k<\Lambda$. However, they are not independent, but related by the splitting Ward identities. Finally, although $w_\Lambda=0$, the corresponding operator is allowed by the symmetries, and will be generated during the flow, and is of order $K$ in our ansatz.

\section{Ward identities} \label{sec_WI} 

\subsection{Splitting symmetry on $\M$}
In flat models the split of the field~$\phi$ into a classical background~$\varphi$ and the corresponding quantum fluctuation~$\xi$ is linear, i.e. $\phi=\varphi + \xi$. This yields a very simple splitting symmetry transformation:  $\varphi \mapsto \tilde{\varphi}=\varphi  + c $, $\xi \mapsto \tilde{\xi}=\xi - c$ where $c$ is a shift. In our case the split is non-linear. To proceed with the background field method we need the transformation rule of the tangent vector~$\xi$ under an infinitesimal small shift~$c$ of the expansion point,
\begin{equation}
  \varphi^{\dot{\lambda}} \mapsto \tilde{\varphi}^{\dot{\lambda}}=e^{c^{\balpha} D_{\balpha}} \varphi^{\dot{\lambda}}= \varphi^{\dot{\lambda}} + c^{\dot{\lambda}} + o(c).
\end{equation}
Here and below the covariant derivative acting on the dotted index is equivalent to the usual partial derivative. Consider the covariant Taylor expansion of the coordinate function
\begin{align}
  \phi^{\dot{\lambda}}&= \varphi^{\dot{\lambda}} + \xi^{\dot{\lambda}} - \sum \limits_{n \geqslant 2} \frac{1}{n!}\xi^{\alpha_1} \dots \xi^{\alpha_n} M^{\dot{\lambda}}_{\alpha_1 \dots \alpha_n} ,\label{0107a}\\
                       M^{\dot{\lambda}}_{\alpha_1 \dots \alpha_n}&=- \frac{1}{n!} \sum \limits_{\pi \in S_n} D_{\pi_1} \dots D_{\pi_2}\varphi^{\dot{\lambda}} ,
\end{align}
where~$S_n$ is the symmetry group on the indices $\alpha_1 \dots \alpha_n$. It is convenient to define the covariant variation of the tangent vector
\begin{equation}
\delta \xi^{\alpha}=D_c \xi^{\alpha}=c^{\gamma}D_{\gamma} \xi^{\alpha}.
\end{equation}
Performing the shift of the expansion point in the Taylor expansion~\eqref{0107a} we obtain
\begin{equation}
 0=\varepsilon^{\dot{\lambda}}-\sum \limits_{n \geqslant 1} \frac{1}{n!}  \varepsilon^{\lambda} \xi^{\alpha_1} \dots \xi^{\alpha_n} M^{\dot{\lambda}}_{\lambda \alpha_1 \dots \alpha_n}  
   - \sum \limits_{n \geqslant 2} \frac{1}{n!} c^{\omega}\xi^{\alpha_1} \dots \xi^{\alpha_n} (D_\omega M^{\dot{\lambda}}_{\alpha_1 \dots \alpha_n} - M^{\dot{\lambda}}_{\omega \alpha_1 \dots \alpha_n}).  \label{0107b}
\end{equation}
where $\varepsilon^{\dot{\lambda}}= c^{\dot{\lambda}} + \delta \xi^{\dot{\lambda}}$. We are looking for the variation~$\delta \xi$ in the form
\begin{equation}
  - \delta \xi^{\dot{\lambda}}=  c^{\dot{\lambda}} + \sum \limits^{\infty}_{m=2} \frac{L^{\dot{\lambda}}_{\omega \beta_1 \dots \beta_m}}{m!}c^{\omega} \xi^{\beta_1} \dots \xi^{\beta_m}. \label{1709a}
\end{equation}
Substitution in Eq.~\eqref{0107b} $\delta \xi$ with the series yields a recurrent relation
\begin{equation}
  L^{\dot{\lambda}}_{\omega \beta_1 \dots \beta_n} = \sum \limits^{n-2}_{m=1 } \frac{n!}{(n-m) ! m !}M^{\dot{\lambda}}_{\beta_1 \dots \beta_{m} \sigma} L^{\sigma}_{\omega \beta_{m+1} \dots \beta_{n}} + M^{\dot{\lambda}}_{\omega \beta_1 \dots \beta_n} - D_\omega M^{\dot{\lambda}}_{\beta_1 \dots \beta_n} .
\end{equation}
We have performed calculation for an arbitrary symmetric manifold, $D_{\sigma} R_{\alpha \lambda \beta \gamma}=0$. Denote by~$\overset{\pi \in S_n}{=}$ the equality under the permutations of the symmetry group~$S_n$. First we turn our attention to the terms on the right hand side which are independent of the unknown coefficients~$L^{\dot{\lambda}}_{\omega  \beta_1 \dots \beta_n}$,
\begin{align}
  M^{\dot{\lambda}}_{\omega \pi_1 \pi_2}- D_\omega M^{\dot{\lambda}}_{\pi_1 \pi_2} &\overset{\pi \in S_2}{=}  \frac{2}{3} R^{\dot{\lambda}}_{\pi_2 \pi_1 \omega},\\
  M^{\dot{\lambda}}_{\omega \pi_1 \pi_2 \pi_3}-D_\omega M^{\dot{\lambda}}_{\pi_1 \pi_2 \pi_3} &\overset{\pi \in S_3}{=} - 2 R^{\sigma}_{\pi_1 \pi_2 \omega} M^{\dot{\lambda}}_{\sigma \pi_3},\\
  M^{\dot{\lambda}}_{\omega \pi_1 \pi_2 \pi_3 \pi_4}- D_\omega M^{\dot{\lambda}}_{\pi_1 \pi_2 \pi_3 \pi_4}&\overset{\pi \in S_4}{=}   4 R^{\sigma}_{\pi_1 \pi_2 \omega} M^{\dot{\lambda}}_{\sigma \pi_3 \pi_4} - \frac{8}{15} R^{\sigma}_{\pi_1 \pi_2 \omega} R^{\dot{\lambda}}_{ \pi_3 \pi_4 \sigma}.  
\end{align}
Then using the recurrent relation we sequentially find the first three coefficients
\begin{equation}
  - \delta \xi^{\alpha}=c^{\alpha} + \frac{1}{3}R^{\alpha}_{\mu \nu \omega} \xi^{\mu} \xi^{\nu} c^{\omega} -  \frac{1}{45} R^\alpha_{\mu \nu \gamma} R^\gamma_{\rho \sigma \omega} \xi^\mu \xi^\nu \xi^\rho \xi^\sigma c^\omega + o(K^2) \label{eq_deltaxi}.
\end{equation}
This geometrical transformation rule implies that given an expansion point~$\varphi$ and the action functional~$S(\phi)$ of the non-linear $\sigma$-model on a symmetric manifold the following identity holds for the covariant Taylor expansion of $S[\vphi,\xi]=S(\phi(\varphi,\xi))$ 
\begin{equation}
  c^\omega\left(D_{\varphi^{\omega}} -   \Big(h^\alpha_{\; \omega} + \frac{1}{3} R^\alpha_{\mu \nu \omega} \xi^{\mu} \xi^{\nu} -  \frac{1}{45} R^\alpha_{\mu \nu \gamma} R^\gamma_{\rho \sigma \omega} \xi^\mu \xi^\nu \xi^\rho \xi^\sigma + o(K^2)\Big) \frac{\delta }{\delta \xi^\alpha}\right)S[\varphi,\xi]=0 
  \label{eq_dS}.
\end{equation}
The directional derivative $c^\omega D_{\varphi^\omega} S[\varphi,\xi]$ corresponds to the parallel transport of~$\xi$ along~$c$ and has to be calculated with the condition $D_c \xi=0$. Since the transformation~\eqref{eq_deltaxi} is independent of the action there is a less laborious  way to obtain it by considering the splitting symmetry of the expansion in question~(see Appendix~\ref{app_exp_action}).

\subsection{Splitting Ward identities}
With these results, we can now derive the corresponding splitting Ward identities for our model, see also \cite{howe88, reuter98, safari16}. For $\varphi \in \M$ and $c,j \in T_\varphi \M$ we denote by $\tilde{\varphi}=\varphi + c$ and by $\tilde{j} \in T_{\tilde{\varphi}}\M$ the parallel transport of~$j$ from~$\varphi$ to~$\tilde{\varphi}$,
\begin{align}
  e^{W_k[\tilde{\varphi} ,\tilde{j}]}&=\int \mathcal{D}_{\tilde{\varphi}}(\tilde{\xi}) e^{-\mathcal{A}_k[\tilde\vphi,\tilde\xi]+ \tilde{j}.\tilde{\xi} },&\mathcal{A}_{k}[\tilde \varphi,\tilde \xi]&=S[\tilde \varphi,\tilde \xi]-\Delta S_k[\tilde \varphi,\tilde \xi].
\end{align}
Since $D_c (\det h(\varphi))=0$ the measure is invariant under the parallel transport,
\begin{equation}
  \mathcal{D}_{\tilde{\varphi}}(\tilde{\xi})= \mathcal{D}_\varphi(\xi).
\end{equation}  
For an infinitesimally small~$c$ this implies
\begin{align}
  e^{W_k[\tilde{\varphi} ,\tilde{j}]}&=\int \mathcal{D}_{\varphi}(\xi) e^{- \mathcal{A}_k[\vphi,\xi] - \delta \mathcal{A}_k[\vphi,\xi] + j \xi },&\delta \mathcal{A}[\varphi,\xi] &=D_c  \mathcal{A}_k[\varphi,\xi].
\end{align}
Then we change the variables $\xi=\xi' + \delta \xi$.  From equation~\eqref{eq_deltaxi} we obtain the Jacobian and the variation of~$U_\varphi$ under this change
\begin{equation}
\begin{split}
\prod \limits_{\balpha}d \xi^\balpha&=\prod \limits_{\balpha}d \tilde\xi^\balpha e^{- \delta_0 (\D -1)\int_x  \xi.c \left( \frac{K}{3}+\frac{K^2}{45} \xi^2\right)} ,\\
U_\Lambda[\varphi,\xi]&=U_\Lambda[\varphi,\tilde\xi] -  \rho_{2,\Lambda} \frac{K(\D -1)}{3}  \int_x \xi.c - \rho_{4,\Lambda} \frac{K^2 (\D -1)}{45} \int_x \xi^2 \xi.c + o(K^2).
\end{split}
\end{equation}
Recall that $\rho^\Lambda_2=\rho^\Lambda_4=\delta_0$. Consequently the measure is invariant also under the variation~$\delta \xi$
\begin{equation}
\mathcal D_\vphi(\xi)=\mathcal D_\vphi(\xi').
\end{equation}
If we did not include the functional~$U_\Lambda$ into the definition of the measure~$\mathcal{D}_\varphi$ we would obtain an anomaly in the Ward identities. Then the splitting identity~\eqref{eq_dS} yields

\begin{equation}
  e^{W_k[\tilde{\varphi} ,\tilde{j}]}=\int \mathcal{D}_{\varphi}(\xi) e^{- \mathcal{A}_k[\vphi,\xi]  + j. \xi - \frac{1}{2}h_{\balpha \bbeta} \xi^\balpha D_c \mathcal{R}_k \xi^\bbeta + (j - \xi \mathcal{R}_{k}) . \delta \xi }.
\end{equation}
To proceed further we introduce an auxiliary source~$\gamma$,
\begin{equation}
  S[\varphi,\xi, \gamma]=S[\varphi, \xi] +  \gamma . \delta \xi\;.
\end{equation}
Thus for the directional derivative we obtain
\begin{align}
  D_c  W_k[\varphi ,j]  &= \Tr\Big(\Big(\frac{\delta W_k[\varphi ,j]}{\delta j} \mathcal{R}_{k} - j\Big)W_{\gamma,k}[\varphi ,j] + \mathcal{R}_{k} \frac{\delta W_{\gamma,k}[\varphi ,j]}{\delta j}\Big)\nonumber \\
                        &\quad - \frac{1}{2} \Tr\Big( \frac{\delta W_k[\varphi ,j]}{\delta j} D_c \mathcal{R}_{k} \frac{\delta W_k[\varphi ,j]}{\delta j} -  D_c \mathcal{R}_{k} \frac{\delta^2 W_k[\varphi ,j]}{\delta j \delta j}\Big),\\
  W_{\gamma,k}[\varphi ,j]&=\frac{\delta W_k[\varphi ,j,\gamma]}{\delta \gamma} \Big|_{\gamma=0}.
\end{align}
Under the parallel transport $D_c j=0$ and $D_c \oxi=0$. Consequently for the directional derivative of the (true) Legendre transform of $W_k[\varphi ,j]$, $\digamma_k[\varphi ,\oxi]=\Gamma_k[\varphi ,\oxi]+\Delta S_k[\vphi,\oxi]$, we have
\begin{align}
  D_c  \digamma_k[\varphi ,\oxi]= -  D_c W_k[\varphi ,j].
\end{align}
Eventually we get the Ward identity
\begin{equation}
\Big(D_c  + \digamma_{  \gamma_\balpha,k}[\varphi,\oxi] \frac{\delta }{\delta \oxi^\balpha}\Big)\Big(\digamma_k[\varphi,\oxi] - \frac{1}{2} h_{\balpha \bbeta} \oxi^\balpha \mathcal{R}_{k} \oxi^\bbeta \Big)=\mathcal{N}_{\varphi k},
\end{equation}
\begin{equation}
\mathcal{N}_{\varphi k}=\Tr \Big(\Big(\frac{\delta \digamma_{  \gamma,k}[\varphi,\oxi]}{\delta \oxi}  \mathcal{R}_{k}+ \frac{1}{2} D_c \mathcal{R}_{k} \Big)(\digamma^{(2)}_k[\varphi,\oxi])^{-1} \Big).
\end{equation}
For the Wetterich effective action the splitting Ward identity has the form \cite{safari16}
\begin{equation}
  \begin{split}
    &D_c \Gamma_k[\varphi,\oxi] + \Gamma_{ \gamma,k}[\varphi ,  \oxi] \frac{\delta  \Gamma_{k}[\varphi,\oxi] }{\delta \oxi^\balpha}=\mathcal{N}_{\varphi k},\\
    &\mathcal{N}_{\varphi k}=\Tr \left(\left(\frac{\delta \Gamma_{ \gamma,k}[ \varphi,\oxi]}{\delta \oxi} \mathcal{R}_{k}+ \frac{1}{2} D_c \mathcal{R}_{k} \right)\left(\Gamma^{(2)}_k[\varphi,\oxi] + \mathcal{R}_{k}\right)^{-1}  \right).
  \end{split}
  \label{eq_WI}
\end{equation}

\subsection{Constraints from the splitting Ward identities to linear order in $K$}
To linear order in $K$, we choose the ansatz for the insertion as follows
\begin{equation}
\Gamma_{\gamma_\alpha,k }[\vphi,\oxi]=  - \zeta_{0,k} c^\alpha -\frac{ \zeta_{2,k}  }{3 } R^\alpha _{\; \mu \nu \omega}\oxi^\mu \oxi^\nu c^\omega+o(K).\label{eq_ansatz_dxi}                               
\end{equation}
This form generalizes Eq.~\eqref{eq_deltaxi} by introducing two coupling constants $\zeta_{0,k} $ and $\zeta_{2,k}$. The ansatz is consistent with the flow equation for $\Gamma_{\gamma_\alpha,k }$ to leading order in $K$.
  
The combination of Eq.~\eqref{eq_ansatz}, Eq.~\eqref{eq_ansatz_dxi} and Eq.~\eqref{eq_WI} gives
  \begin{align}
  \zeta_{0,k}\frac{t_{n,k}}{t_{n+1,k}}&=1+O(K),&\zeta_{0,k} \zeta_{2,k}&=1+O(K),&t_{2,k}\upsilon_k&=1+O(K),&t_{2,k}w_k&=O(K).
  \label{eq_constraint_WI}
\end{align}
  
  Analysing the Ward identity Eq.~\eqref{eq_WI}, one finds that $\rho_{2,k}$ is not an independent variable, but obeys
  \begin{equation}
  -\rho_{2,k}\zeta_{0,k}\frac{\delta U^{(2)}}{\delta \oxi^\balpha}c^\balpha=\Tr\left(\mathcal R_k \frac{\delta \Gamma_{\gamma,k}}{\delta \oxi}\left(\Gamma^{(2)}_k+\mathcal R_k\right)^{-1}\right) + o(K).
\label{eq_WIrho2}
\end{equation}
On the right hand side one only keeps the local term $  \oxi^\balpha c_\balpha$ to leading order in $K$.

\section{FRG flow equations}\label{sec_flow}

\subsection{Method and notations }

In this section we compute the flow equations of the various coupling constants to linear order in $K$ using our ansatz Eq.~\eqref{eq_ansatz}. For conciseness we use the following notations
\begin{align}
  \overline\Gamma^{(n)}_{\balpha_1 \dots \balpha_n,k}&=\frac{\delta^n \Gamma_k[\vphi,\oxi]}{ \delta \oxi^{\balpha_1} \dots \delta \oxi^{\balpha_n} }\Big|_{\oxi=0}\\
  \overline G_k &=\left(\overline\Gamma^{(2)}_k+\mathcal R_k\right)^{-1},
\end{align}
where
\begin{equation}
\overline\Gamma^{(2)}_k = h\frac1{t_{2,k}}\left(- D^2+m_k^2\right)+\Sigma,
\end{equation}
with $\Sigma_{\alpha \beta}= \upsilon_k E_{\alpha \beta} +  w_k E^\gamma_{\; \gamma} h_{\alpha \beta}$ and $m_k^2=\frac{K(\D-1) }{3}\rho_{2,k} t_{2,k}$. We will see below that $t_{2,k}$ is of order $K^{-1}$ at the fixed point. Consequently $t_{2,k}\Sigma$ is of order $K$, while $m_k^2$ will be of order one. We choose the regulator function of the form
\begin{equation}
\mathcal{R}_{\balpha \bbeta,k}[\vphi]=\frac1{t_{2,k}}h_{\alpha\beta}R_k(-D^2),
\label{eq_reg}
\end{equation}
with $R_k(\omega)=(k^2-\omega)\theta(k^2-\omega)$.

Then, for a sufficiently small $K$, we assume the existence of the Neumann series 
\begin{align}
  \overline G_k&=t_{2,k}  \G h^{-1} \sum \limits^\infty_{n=0} \left( -t_{2,k}\Sigma\G  h^{-1} \right)^{n}
\end{align}
where
\begin{align}
  \G^{-1}&=- D^2 +  R_k + m_k^2 .\label{2206a}
\end{align}
Note that we do not expand $\overline \G$  in powers of $m_k^2$. 

To compute the trace, we use the heat kernel method~\cite{avramidi15, vassilevich2003}, that we outline briefly.
The spectral decomposition of a integral kernel~$f$ reads
\begin{equation}
  f^{\alpha  \beta }(x,y)= \sum \limits_{\omega \in \sigma(-\Delta)} \hat{f}(\omega) \psi^{\alpha \, \omega }(x) \left(\psi^{\beta \, \omega}(y)\right)^*=\int \limits^\infty_0 ds \,(\mathcal{L}^{-1} \hat{f})(s) \mathcal{K}^{\alpha \beta}_{ x y}(s), \label{2304a}
\end{equation}
where $-D^2\psi^{ \omega }=\omega \psi^{\omega }$.
The heat kernel~$\mathcal{K}$ satisfies the heat equation $(\partial_s - D^2)\mathcal{K}_{xy}(s)=0$. For $k^2> \|\partial\vphi\|^2_\infty$ the inverse Laplace transform $(\mathcal{L}^{-1} \hat{f})(s)$ is small for all large values of time, i.e. such that $s \|\partial\vphi\|^2_\infty>1$. Consequently one can substitute the heat kernel with an asymptotic expansion at small time
\begin{equation}
  \mathcal{K}_{xy}(s)=(4 \pi s)^{-\frac{d}{2}}e^{-\frac{(x-y)^2}{4s }} \sum \limits^\infty_{m=1}\frac{(-1)^m}{m!} b_{m}(x,y) \, s^m \, .\label{2304b}
\end{equation}
At the coincidence limit $y \to x$ the leading heat kernel coefficients are~\cite{zanusso}
\begin{align}
  b_0&=1,&D_{x_i} b_0 &=0,\\
  b_1&=0, &D_{x_i}   b_1&= \frac{D_k \Omega_{k i}}{6} ,&\Omega_{\alpha \beta \; k i}&=-K\Pi_{\alpha \beta \lambda \gamma} \partial_k \vphi^\lambda \partial_i \vphi^\gamma.
\end{align}
For a spectral density $\hat{f}(\omega)$, we introduce 
\begin{equation}
    Q_{\frac{d}{2} - m}[\hat{f}]=  \int \limits^\infty_0 ds \,(\mathcal{L}^{-1} \hat{f})(s) \frac{s^m}{(4 \pi s)^{ \frac{d}{2}}},
  \end{equation}
  that for $d>2m $  is easier to calculate in the spectral representation
\begin{equation}
Q_{\frac{d}{2} - m}[\hat{f}]=  \frac{1}{(4 \pi)^{\frac{d}{2}} \Gamma(\frac{d}{2} - m)} \int \limits^\infty_0 d \omega \, \hat{f}(\omega)\, \omega^{\frac{d}{2} - m -1}.  
\end{equation}
In particular, for 
\begin{equation}
  \h= \G \left( \partial_k R_k-\frac{\partial_k t_{2,k}}{t_{2,k}}R_k\right) \G,
\end{equation}
one finds
\begin{equation}
  Q_{\frac{d}{2} - m}[\hat{\h}]=  \frac{2 k^{d +1 - 2m}}{(4 \pi )^{\frac{d}{2}} \Gamma(\frac{d}{2} + 1 - m )(k^2 + m_k^2)^{2}} \left(1 -  \frac{k\partial_k{t}_{2,k} }{t_{2,k} (d - 2m + 2)  }  \right).
\end{equation}

\subsection{Flow of the effective action}

The flow equation of  $\bar\Gamma_k$ is 
\begin{align}
  \partial_k\bar\Gamma_k= \frac{1}{2} \Tr \left( \partial_k{\mathcal{R}_k} \overline G_k\right). \label{1106a}
\end{align}
To leading order in $K$, we obtain
\begin{equation}
 \frac{\partial_k{t_{0,k}}}{t^2_{0,k}}= -t_{2,k} (\upsilon_k + \D w_k) K (\D -1) Q_{\frac{d}{2}}[\hat{\h}] . \label{eq_flow_t0}
\end{equation}
The flow of the one-point function $\overline \Gamma^{(1)}_k$ reads
\begin{equation}
  \partial_k\overline \Gamma^{(1)}_{ \balpha,k}= -\frac{1}{2} \Tr \left( \partial_k{\mathcal{R}_k}\overline G_k \overline\Gamma^{(3)}_{\balpha,k} \overline G_k \right).
\end{equation}
Using the ansatz in Eqs.~\eqref{eq_ansatz} and \eqref{eq_VU} we have
\begin{align}
  \overline\Gamma^{(3) }_{\bmu \bbeta \balpha}&=  \sum \limits_{\pi \in S_3} \int_x \frac{-2K}{3 t_{3,k}} \Pi_{\pi_\balpha \blambda \pi_\bbeta \bgamma}(\varphi _x) \partial_i \varphi^{\blambda} {D_i}^{\bgamma}_{\pi_\bmu},\\
  \overline\Gamma^{(4) }_{ \bnu \bmu \bbeta \balpha}&= \sum \limits_{\pi \in S_4} \int_x   \frac{-K}{3! t_{4,k}} \left( \Pi_{\pi_\balpha \blambda \pi_\bbeta \bgamma}(\varphi_x)   {D_i}^{\bgamma}_{\pi_\bmu}  {D_i}^{\blambda}_{\pi_\bnu} +  h_{\pi_\balpha \pi_\bbeta}(\varphi_x) E_{\pi_\bmu \pi_\bnu}(\varphi_x) \right)\nonumber \\&\quad  + \frac{K^2 (\D-1) \rho_{4,k}}{180} h_{\pi_\balpha \pi_\bbeta}(\varphi_x)  h_{\pi_\bmu \pi_\bnu}(\varphi_x).
\end{align}
Here $h_{\mu z \, \nu \bar{z}}(\varphi _x)=h_{\mu  \nu }(\varphi_x) \delta_{x z} \delta_{x \bar{z}}$ and ${D_i}^{\gamma x}_{\mu z}= \partial_{x_i}  \delta_{xz} h^{\gamma}_{\; \mu} + \Gamma^{\gamma}_{\mu \sigma} \partial_i \varphi^\sigma_x \delta_{xz}$, i.e. the covariant derivative with respect to the upper index.

To leading order in~$K$ the equation has the form
\begin{equation}
   \partial_k\overline \Gamma^{(1)}_{ \balpha,k}=-\frac{t_{2,k}}{2} \Tr \left(\h \overline\Gamma^{(3)}_{\balpha,k}\right) + o(K).
\end{equation}
This gives
\begin{equation}
  \frac{\partial_k{t_{1,k}}}{t^2_{1,k}} = -2\frac{ t_{2,k}}{ t_{3,k}} \frac{K (\D -1)}{3}  Q_{\frac{d}{2}}[\hat{\h}] \,. \label{eq_flow_t1}
\end{equation}
Finally, the flow of the two-point function reads
\begin{align}
   \partial_k\overline\Gamma^{(2)}_{ \balpha \bbeta,k}&=\frac{1}{2} \sum \limits_{\pi \in S_2} \Tr \left(\partial_k{\mathcal{R}_k}\overline G_k \overline\Gamma^{(3)}_{\pi_{\bbeta},k}  \overline G_k \overline\Gamma^{(3)}_{\pi_{\balpha},k}\overline G_k  \right)\nonumber\\
 &\quad -\frac{1}{2} \Tr \left(\partial_k{\mathcal{R}_k}\overline G_k\overline\Gamma^{(4) }_{ \balpha \bbeta,k} \overline G_k  \right), \label{1603a}
\end{align}
where $S_2$ is the symmetry group on two indices $\balpha$, $\bbeta$. To leading order in $K$, the equation is
\begin{equation}
 \partial_k\overline\Gamma^{(2)}_{ \bnu \bar{\bnu},k } =   -  \frac{t_{2,k}}{2} \Tr \left( \h  \overline\Gamma^{(4)}_{ \bnu \bar{\bnu} ,k } \right) + o(K)  . \label{1204a}
\end{equation}
 It is convenient to write this flow using an auxiliary generating functional
 \begin{equation}
    F(\xi,\bar{\xi})=\frac{t_{2,k}}{2} \Tr \left(  \h  \overline \Gamma^{(4)}_{\bnu \bmu,k }  \right) \xi^\bmu \bar{\xi}^\bnu,
\end{equation}
which reads after expansion to leading order in $K$ and to second order in derivatives
\begin{equation}
  F= \int_x    \xi_\alpha (- (\D -1 ) r_0  D^2  + l_1)\bar{\xi}^\alpha   + (r_{0} h_{\alpha \beta} E^{\gamma}_{\; \gamma}  + (\D + 4) r_0 E_{\alpha \beta})\xi^\alpha \bar{\xi}^\beta    + o(K).
  \label{eq_flowG2}
\end{equation}
The auxiliary constants are as follows 
\begin{align}
r_i&=-\frac{K t_{2,k}}{3 t_{4,k}}Q_{\frac{d}{2}+i}[\hat{\h}],&l_1 &=    \frac{d}{2} (\D -1 ) r_{1} +  \frac{ K^2(\D+2)(\D -1 )\rho_{4,k} t_{2,k}}{45} Q_{\frac{d}{2}}[\hat{\h}].
\end{align}
Eq.~\eqref{1204a} yields the evolution equation for the constant~$t_{2,k}$
\begin{equation}
  \frac{\partial_k t_{2,k}}{t^2_{2,k}} = -\frac{ t_{2,k}}{ t_{4,k}} \frac{K (\D -1)}{3} Q_{\frac{d}{2}}[\hat{\h}] \,. \label{eq_flow_t2}
\end{equation}
From Eq.~\eqref{eq_flowG2}, one can also obtain the flow of $\upsilon_k$, $w_k$ and $\rho_{2,k}$, although these will not be needed as they are fixed by the splitting Ward identity.

For flat models one makes the usual substitution $t_{i,k}=t_k$ where $t_k$ is a unique renormalized coupling constant. This makes possible to retain only the evolution equations for 1PI vertex functions~$\bar \Gamma^{(n)}$ with $n < 2$. For non-linear $\sigma$-models the substitution $t_{i,k}=t_k$ would give incorrect flow equations (e.g. comparing Eqs.~\eqref{eq_flow_t1} and \eqref{eq_flow_t2}).

\subsection{Flow of $\Gamma_{\gamma,k}$ }
The flow equation of $\Gamma_{\gamma,k}$ reads
\begin{equation}
 \partial_k \Gamma_{\gamma,k}= -\frac{1}{2} \Tr \left( \partial_k \mathcal R_k \overline G_k \overline\Gamma^{(2)}_{\gamma,k} \overline G_k \right),\\
\end{equation}
that at leading order in $K$ takes the form
\begin{equation}
\partial_k \Gamma_{\gamma_\alpha,k}=-\frac{t_{2,k}}{2} \Tr (\h \overline\Gamma^{(2)}_{\gamma_\alpha})=   -\frac{ K(\D -1)  }{3} t_{2,k}\zeta_{2,k}  Q_{\frac{d}{2}}[\hat{\h}]c^\alpha + O(K).
\end{equation}
Consequently we have 
\begin{equation}
   \partial_k\zeta_{0,k}= \frac{K(\D -1) }{3}   \zeta_{2,k}  t_{2,k}Q_{\frac{d}{2}}[\hat{\h}]. \label{eq_flow_zeta0}
\end{equation}
Finally,  for our choice of regulator function Eq.~\eqref{eq_reg}, the Ward identity Eq.~\eqref{eq_WIrho2} written in terms of $m^2_k$ reads
\begin{equation}
  m^2_k=\frac{\zeta_{2,k}t_{2,k}}{\zeta_{0,k}} \frac{K(\D-1)}{3}Q_{\frac d2}[\hat R \hat \G], 
  \label{eq_WIm2}
\end{equation}
with
\begin{equation}
  Q_{\frac d2}[\hat R \hat \G] =   \frac{ k^{d +2}}{(4 \pi )^{\frac{d}{2}}\Gamma(\frac{d}{2} + 2 )(k^2 + m_k^2) }.
  \end{equation}

\section{$\beta$-functions and fixed point analysis}\label{sec_betafunc}

Using Ward identities Eq.~\eqref{eq_constraint_WI} the flow of $t_{0,k}$ Eq.~\eqref{eq_flow_t0} can be written to leading order in $K$ as
\begin{equation}
  \frac{\partial_k t_{0,k}}{t^2_{0,k}}= -K (\D -1) Q_{\frac{d}{2}}[\hat{\h}] ,
\end{equation}
while that of $t_{1,k}$, $t_{2,k}$ and $\zeta_{0,k}$ takes the simple form
\begin{align}
   \frac{\partial_k t_{1,k}}{t_{1,k}} &= \frac23\frac{\partial_k t_{0,k}}{t_{0,k}}, & \frac{\partial_k t_{2,k}}{t_{2,k}} &= \frac13\frac{\partial_k t_{0,k}}{t_{0,k}},& \frac{\partial_k\zeta_{0,k}}{\zeta_{0,k}} &=- \frac13\frac{\partial_k t_{0,k}}{t_{0,k}}. \label{2306a}
\end{align}
Furthermore, recalling that $m_k^2=\frac{K(\D-1) }{3}\rho_{2,k} t_{2,k}$ and using Eqs.~\eqref{eq_constraint_WI} and \eqref{eq_WIm2}, one finds $m^2_k$ as a function of $t_{0,k}$,
\begin{equation}
 m^2_k=\frac{2s_d}{3(d+2)}\frac{ k^{d +2}t_{0,k}}{k^2 + m_k^2 },
  \label{eq_m2}
\end{equation}
with $s_d=\frac{K (\D -1)}{(4 \pi)^{\frac{d}{2}} \Gamma(\frac{d}{2} +1 ) }$.
Since $Q_{\frac{d}{2}}[\hat{\h}]$ depends on $\frac{\partial_k t_{2,k}}{t_{2,k}}$ and $m^2_k$, this allows to write the flow equation of $t_{0,k}$ in terms of $t_{0,k}$ only,
\begin{equation}
k\partial_k t_{0,k}=-\frac{2 s_d k^{d+2}t_{0,k}^2}{(k^2+m_k^2)^2}\left(1-\frac{k\partial_k t_{0,k}}{3(d+2) t_{0,k}}\right).
\end{equation}

To analyse the flow equations, it is convenient to introduce dimensionless variables $\tilde t_{0,k}=k^{d-2}t_{0,k}$ and $\tilde m^2_k=k^{-2}m^2_k$. For the latter, by keeping only the positive root when solving Eq.~\eqref{eq_m2}, we obtain
\begin{equation}
\tilde m_k^2=\frac{\sqrt{1+\frac{8 s_d \tilde t_{0,k}}{3(d+2)}}-1}{2}. \label{2206b}
\end{equation}
Defining the $\beta$-function, $\beta_0=k\partial_k \tilde t_{0,k}$, our final result is
\begin{equation}
\beta_0 = (d-2) \tilde t_{0,k} - \frac{4 s_d \tilde t_{0,k}^2}{1+\sqrt{1+\frac{8 s_d \tilde t_{0,k}}{3(d+2)}}}.
\end{equation}
A fixed point is a scale independent solution, i.e. $\beta_0=0$. There are two fixed points associated to this $\beta$-function, the trivial fixed point $\tilde t^*_{0,k}=0$, which is attractive in the infrared and corresponds to the low-temperature phase, and a non-trivial fixed point
\begin{equation}
\tilde t_0^*=\frac{2(d+1)(d-2)}{3(d+2) s_d},
\end{equation}
for $d>2$ and if $s_d$ is positive. For $K>0$, the model is the usual $O(\D)$ NLsM, while for $K<0$, the fixed point is physical in the formal limit $D<1$, and in particular for $\D=0$. Expanding the $\beta$-function at~$\tilde{t}^*_0$ we obtain the linearised equation
\begin{align}
  k \partial_k \tilde{t}_{0,k}&= -\nu^{-1}(\tilde{t}_{0,k} - \tilde{t}^*_{0}) + o(\tilde{t}_{0,k} - \tilde{t}^*_{0}),&\nu^{-1}&=(d-2)\left(1 - \frac{d-2}{5d +2}\right),
\end{align}
with $\nu^{-1}$ the critical exponent governing the divergence of the correlation length close to criticality. In particular for $d=3$ we have $\nu^{-1}=  16/17$, for all $\D$ and $K$ such that $s_3>0$. Clearly the fixed point is repulsive. For $d=2+\epsilon$, $\epsilon\to 0$, we recover the standard one-loop result $\nu^{-1}=\epsilon$ \cite{Brezin1976,Houghton1980}. At the fixed point the Ward identities~\eqref{2206b},~\eqref{2306a} give
\begin{align}
  \tilde{m}^2_k &= \frac{d -2}{3(d+2)},&\rho_2&=\frac{3 \Lambda^d (k \Lambda^{-1})^{2 + \frac{d-2}{3}} }{2  (d+1) (4 \pi)^{\frac{d}{2}} \Gamma(\frac{d}{2} +1)} ,\\
  \zeta^{-1}_2=\zeta_0&= (k \Lambda^{-1} )^{\frac{d-2}{3}},&t_{2,k}&=\tilde{t}^*_0 \Lambda^{2-d} (k \Lambda^{-1})^{\frac{2 - d}{3}}.
\end{align}

\section{Discussion and conclusion \label{sec_concl}}

We have computed the FRG flow equation of NLSM with constant curvature using the background field method, to lowest order in the derivative expansion. In order to implement consistently the splitting Ward identities induced by the background field reparametrization invariance, we have also performed a formal expansion in the curvature, keeping terms to lowest order in $K$. The beta functions we have obtained are different from those of the previous studies using the same method \cite{percacci, flore}, corresponding to different critical exponents (if one stays at the same order of the derivative expansion). Let us comment on the main difference between these works and ours.

In \cite{percacci}, we note that the ``mass'' term induced by the measure is neglected, and that all the $t_{i,k}$ are assumed to be identical, i.e. $t_{i,k}=t_{0,k}$, in our notations. This is obviously not consistent with the splitting Ward identities derived above. In \cite{flore}, a ``wavefunction renormalization'' is introduced for the fluctuating field, corresponding here to $t_{0,k}/t_{2,k}$, as well as a mass term. No connection with the splitting Ward identities is made, and the mass has an independent flow, whereas we have shown that it is fixed by the Ward identities. Therefore, their flow equations at the lowest order in the derivative expansion are different from ours. 
It has been noted in \cite{flore} that if one includes all coupling constants at the next order of the derivative expansion, the non-trivial fixed point disappears. One could hope that using an ansatz that obeys the Ward identities to second order in $K$ will cure this problem. 

One aspect which is identical in our work and \cite{percacci} is that a non-trivial fixed point is found in all dimensions $d>2$, which, if confirmed, implies that there is no upper critical dimension. While this is expected for non-compact NLSM, as discussed in the introduction, the fact that we find the same result for the $O(N)$ NLSM questions the validity of the approach. Indeed, on the lattice, the $O(N)$ NLSM corresponds to a $O(N)$ spin model, for which there is no doubt that the upper critical dimension is $d_c=4$. If, and how, the present method is able to recover this result is still an open question. (We note in passing that a lattice FRG approach of the $O(N)$ NLSM, not using the background field method but taking the non-linear constraint into account exactly, does not suffer from this problem. Indeed, the flow equations are formally the same than that of the corresponding linear sigma model, and only the initial condition is different, which does not affect the fixed point properties \cite{Machado2010}.) It has been argued that the $2+\epsilon$ expansion of the $O(N)$ NLSM does not describe the Wilson-Fisher fixed point at $\epsilon=1$, as it cannot capture the topological excitations that drive the transition, e.g. the hedgehogs excitations for $N=3$ \cite{Nahum2015}. It could well be that the background field method, even supplemented with a functional RG approach, is incapable to capture the correct physics far from $d=2$. 
We hope that the expansion to the next order in derivatives and curvature will help to answer these questions.

\acknowledgements
We thank R. Percacci and A. Codello for correspondence about their work and discussions. AR thanks D. Mouhanna for insightful discussions on the NLSM, as well I. Balog for very useful discussions on the $SO(1,N)$ model. AE is greatful to B. Arras for giving the opportunity to do this work. For all tensor calculus we have used Cadabra~\cite{cadabra2}. This is an extremely lightweight, latex friendly and completely free software tool. Needless to say how easy tensor algebra has nowadays become.
This work was supported by Agence Nationale de la Recherche through Research Grants QRITiC I-SITE ULNE/ ANR-16-IDEX-0004 ULNE, the Labex CEMPI Grant No.ANR-11-LABX-0007-01, the Programme Investissements d'Avenir ANR-11-IDEX-0002-02, reference ANR-10-LABX-0037-NEXT and the Ministry of Higher Education and Research, Hauts-de-France Council and European Regional Development Fund (ERDF) through the Contrat de Projets \'Etat-Region (CPER Photonics for Society, P4S).

\begin{appendix}

\section{Covariant expansion of the action}\label{app_action}
There is a variety of sources where the reader can find the covariant expansion of the NLSM, see e.g. ~\cite{mukhi81,howe88,ket2000},
\begin{align}
  \int \limits_{\mathbb{R}^d} \frac{(\partial \phi)^2}{2}&= \int \limits_{\mathbb{R}^d} \left( \frac{(\partial \varphi)^2}{2} -  \xi_\gamma D_i \partial_i \varphi^\gamma    + \frac{\xi^{\alpha }(C^{-1}_{\alpha \beta} + E_{\alpha \beta})\xi^{\beta }}{2} \right)  + \sum \limits^5_{n=3} V^{(n)}[\varphi,\xi] +o(\xi^5),\label{2007b}\\
  C^{-1}_{\alpha \beta} &=  h_{\alpha \beta}(- D^2),\quad E_{\alpha \beta}=-R_{\lambda \alpha \gamma \beta} \partial_i \varphi^\lambda \partial_i \varphi^\gamma  =- K \Pi_{\alpha \lambda \beta \gamma}  \partial_i \varphi^\lambda \partial_i \varphi^\gamma ,\\
  V^{(3)}[\varphi,\xi] &= - \frac{2}{3 } R_{  \bsigma \balpha \bgamma  \bbeta}  \xi^\balpha \xi^\bbeta \partial_i \varphi^\bsigma D_i \xi^\bgamma  = \frac{-2K}{3 } \int \limits_{\mathbb{R}^d}  \xi^\alpha \xi^\beta   \Pi_{\alpha \lambda \beta \gamma} \partial_i \varphi^\lambda D_i \xi^\gamma,\\
  V^{(4)}[\varphi,\xi] &= - \frac{1}{3!} R_{\bgamma \balpha \bsigma \bbeta } \xi^\balpha \xi^\bbeta \left( D_i \xi^\bgamma D_i \xi^\bsigma  - R^{\bgamma}_{\balpha' \bsigma' \bbeta'} \xi^{\balpha'} \xi^{\bbeta'} \partial_i \varphi^\bsigma \partial_i \varphi^{\bsigma'}\right)\nonumber\\
        &= \frac{-K}{3!} \int  \limits_{\mathbb{R}^d} \xi^\alpha \xi^\beta  \left(\Pi_{\alpha \lambda \beta \gamma}   D_i \xi^\gamma D_i \xi^\lambda  +  \xi^2 E_{\alpha \beta} \right),\\
  V^{(5)}[\varphi,\xi] &=\frac{2}{15}  \int \limits_{\mathbb{R}^d}  R_{\mu \alpha \beta \lambda}  R^\lambda_{\rho \sigma \nu} \xi^{\alpha} \xi^{\beta} \xi^{\rho} \xi^{\sigma} D_i \xi^{\nu} \partial_i \varphi^{\mu}= \frac{2 K^2}{15}  \int \limits_{\mathbb{R}^d}  \xi^2 \Pi_{\alpha \lambda \beta \gamma}\xi^{\alpha} \xi^{\beta}  D_i \xi^{\lambda}   \partial_i \varphi^{\gamma},
\end{align}
where we have factored out the factor $t^{-1}$ in the action.

\section{Splitting symmetry of the action} \label{app_exp_action}
We would like to give a simple method to obtain the symmetry transformation in Eq.~\eqref{eq_deltaxi}. Indeed the covariant Taylor expansion of the action Eq.~\eqref{2007b} is independent of the point~$\varphi$. To proceed we have to retain in the expansion all terms quadratic in the Riemann tensor. The directional derivative vanishes at the first and third orders in~$\xi$ iff~$L^{\dot{\alpha}}_{\omega \beta}=L^{\dot{\alpha}}_{\omega \beta_1 \beta_2 \beta_3}=0$. The definition of~$L^{\dot{\alpha}}_{\omega \beta_1 \dots \beta_m}$ is given in~\eqref{1709a}. Then for $m$ an even integer we put
\begin{equation}
  L_{\dot{\alpha} \omega \beta_1 \beta_2 \dots \beta_{m-1} \beta_m} = \sum \limits_{\pi \in S_m} \left( a_{m  1}\, h_{\dot{\alpha} \omega } h_{\pi_{\beta_1} \pi_{\beta_2}}\dots h_{\pi_{\beta_{m-1}} \pi_{\beta_m}}  + a_{m  2} \, h_{\dot{\alpha} \pi_{\beta_1}} h_{\omega  \pi_{\beta_2}} \dots h_{\beta_{m-1}  \pi_{\beta_m}} \right).
\end{equation}
The derivative vanishes at the second and fourth orders in~$\xi$ iff
\begin{align}
a_{21}&=-\frac{ K}{3},&a_{22}&=\frac{K}{3},&a_{41}&=-\frac{K^2}{45},& a_{42}&=\frac{K^2}{45}.
\end{align}
Once again this yields the symmetry transformation given in \eqref{eq_deltaxi}.

\section{Wilson--Polchinski equation} \label{2107c}
Most equations of this appendix are complementary to those of the main text. However we believe they are likely useful for the reader. Let $C_{k \Lambda}$ be a regularized propagator such that
  \begin{align}
  C_{\Lambda \Lambda}&= 0,& \lim \limits_{\substack{k \to 0\\ \Lambda \to \infty}}  C^{-1}_{k \Lambda} &=  - D^{2}.
  \end{align}
  For $\forall j \in \mathcal{D}(\mathbb{R}^d ,T_{\varphi} \M)$  one can write the partition functional in the form
  \begin{equation}
    Z_{k \Lambda}[\varphi,j]= e^{-\frac{1}{2} \int ( \partial \varphi)^2 - \Gamma_{1,k \Lambda}[\varphi ] + \uw_{k \Lambda}[\varphi,j +  D_i \partial_i \varphi]}.
  \end{equation}
Here $\Gamma_{1, k \Lambda}$ is a normalization coefficient,
\begin{equation}
  \Gamma_{1,k \Lambda}[\varphi]=\frac{1}{2} \Tr \log ( (-\partial^2_{\Lambda})^{-1} \, h^{-1}C^{-1}_{k \Lambda}).\label{507a}
\end{equation}
The generating functional of connected Schwinger functions~$\uw_{k \Lambda}[\varphi,j]$ is
\begin{align}
   e^{\uw_{k \Lambda}[\varphi,j]}&=\int d \mu_{k \Lambda}(\xi)  e^{-  L_{\Lambda}[\varphi,\xi] +   \xi^\balpha j_\balpha},
\end{align}
  where $d \mu_{k \Lambda}$ is a Gaussian measure on a finite dimensional Borel cylinder set~\cite{gljf87},
  \begin{equation}
    d \mu_{k \Lambda}(\xi)=  \sqrt{\det C^{-1}_{k \Lambda}} \prod \limits_{\balpha} \frac{d \xi^\balpha}{\sqrt{2 \pi }}\, e^{-\frac{1}{2 }  \xi^\balpha C^{-1}_{k \Lambda \,\balpha \bbeta} \xi^\bbeta}.
   \end{equation} 
At 1-loop the bare reduced action is (see App.~\ref{app_action})
\begin{align}  
    L_{\Lambda}[\varphi,\xi]&=\left(\frac{1}{t_{0,\Lambda}} -1 \right) h_{\balpha \bbeta}\partial \varphi^\balpha \partial \varphi^\bbeta   + \left(1-\frac{1}{t_{1,\Lambda}} \right) \xi_\balpha D_i \partial_i \varphi^\balpha +  \frac{1}{2} \left(\frac{1}{t_{2,\Lambda}} -1\right)   \xi^\balpha C^{-1}_{0 \Lambda \, \balpha \bbeta} \xi^\bbeta  \nonumber \\
                                    &\quad + \frac{\upsilon_{\Lambda}}{2}E_{\balpha \bbeta} \xi^{\balpha} \xi^{\bbeta}  + \frac{1}{t_{3,\Lambda}}V^{(3)}[\varphi,\xi] + \frac{1}{t_{4,\Lambda}} V^{(4)}[\varphi,\xi] \nonumber\\
  &\quad + \rho_{2,\Lambda} U^{(2)}[\varphi,\xi] + \rho_{4,\Lambda} U^{(4)}[\varphi,\xi]+ o(\xi^4).
\end{align}
The usual way to give a meaningful interpretation of~$\Gamma_{1,k \Lambda}$ is to consider a stationary point of the free energy, 
  \begin{equation}
    \frac{\delta W_{0 \Lambda}[\varphi,j]}{\delta j}=0.
  \end{equation}
Using convexity of the effective action one shows that at this point the normalization coefficient~\eqref{507a} is the effective action at 1-loop~\cite{jackiw74}. 

It is convenient to define the reduced effective action~$\uf_{k \Lambda}[\varphi,\oxi]$
\begin{equation}
  \mathcal{L}(\uw_{k \Lambda}[\varphi,\cdot])(\oxi)=\frac{1}{2} \oxi^\balpha C^{-1}_{k \Lambda \, \balpha \bbeta} \oxi^\bbeta + \uf_{k \Lambda}[\varphi,\oxi],
\end{equation}
where $\mathcal{L}( \cdot)$ is the Legendre transform. Then the Wilson--Polchinski equation~\cite{pol84,mor94,bdm93} is
\begin{align}
  \partial_k \uf_{k  \Lambda}[\varphi,\oxi]&= \frac{1}{2} \Tr \left( \partial_k C_{k \Lambda} \, \uf^{(2)}_{k  \Lambda}[\varphi,\oxi](1 + C_{k  \Lambda} \uf^{(2)}_{k  \Lambda}[\varphi,\oxi])^{-1} \right),\label{0306a}\\
    \uf_{\Lambda  \Lambda}[\varphi,\oxi]&=L_{\Lambda}[\varphi, \oxi]. \label{507b}               
\end{align}
Substituting $C^{-1}_{k \Lambda}=C^{-1}_{0 \Lambda} + \mathcal{R}$ into the Wetterich effective action~\eqref{2107b} we obtain
\begin{equation}
  \Gamma_{k \Lambda}[\varphi,\oxi]=\frac{1}{2}h_{\balpha \bbeta} \partial_i \varphi^\balpha \partial_i \varphi^\bbeta  - \oxi^\balpha D_i \partial_i \varphi_{ \balpha} + \frac{1}{2} \oxi^\balpha C^{-1}_{0 \Lambda \, \balpha \bbeta} \oxi^\bbeta  + \Gamma_{1,k \Lambda}[\varphi,\oxi] +\uf_{k \Lambda}[\varphi,\oxi].
\end{equation}
It follows that this action satisfies the following boundary condition  
\begin{equation}
  \lim \limits_{k \to \Lambda} \left(\Gamma_{k  \Lambda}[\varphi,\oxi] - \frac{1}{2} \Tr \log (-\partial^{2}_{\Lambda})^{-1}  h^{-1} C^{-1}_{k  \Lambda} \right) = S_{\Lambda}[\varphi,\oxi] + U_{\Lambda}[\varphi,\oxi]. 
\end{equation}
On the right hand side we used $t_{i,\Lambda}=t$, see Eq.~\eqref{eq_IC}.
\end{appendix}
\bibliography{bibli_nlsm_jphys}
\end{document}